# Solvothermal Reduction of Chemically Exfoliated Graphene Sheets

Hailiang Wang, Joshua Tucker Robinson, Xiaolin Li, and Hongjie Dai*

*Department of Chemistry and Laboratory for Advanced Materials, Stanford University, Stanford, CA 94305, USA*

Graphene has attracted much attention due to its interesting properties and potential applications.[1-3] Chemical exfoliation methods have been developed to make graphene recently,[4,5] aimed at large-scale assembly[6] and applications such as composites[7] and Li-ion batteries.[8] Although efficient, the chemical exfoliation methods involve oxidation of graphene and introduce defects in the as-made sheets. Hydrazine reduction at ≤100℃ has shown to partially restore the structure and conductance of graphite oxide (GO).[4,9-11] However, the reduced GO still shows strong defect peaks in Raman spectra with higher resistivity than pristine graphene by 2 to 3 orders of magnitude.[4,9-11] It is important to produce much less defective graphene sheets (GS) than GO, and develop more effective graphene reduction.

Recently, we reported a mild exfoliation-reintercalation-expansion method to form high-quality GS with higher conductivity and lower oxidation degree than GO.[5] Here, we present a 180℃ solvothermal reduction method for our GS and GO. The solvothermal reduction is more effective than the earlier reduction methods in lowering the oxygen and defect levels in GS, increasing the graphene domains, and bringing the conductivity of GS close to pristine graphene. The reduced GS possess the highest degree of pristinity among chemically derived graphene.

GS were made from natural graphite flakes, intercalated by oleum and tetrabutylammonium cations, and suspended into N, N- dimethylformamide (DMF).[5,6] Solvothermal reduction was carried out in DMF at 180℃, using hydrazine monohydrate as the reducing agent. The GS (average size ~ 300nm on the side) remained well-dispersed in DMF after reduction. The homogeneous suspension contained mostly single sheets (Fig.1) observed by atomic force microscopy (AFM) on $SiO_2$. The apparent height of the GS were about 0.8-1.0nm, suggesting single-layer GS.

We also reduced our GS at room temperature and 80℃ with hydrazine monohydrate. After reductions, the Raman D/G intensity ratios (D peak is a defect peak due to inter-valley scattering[12]) were higher than as-made GS (Fig.2a,b), indicating an increase in the number of smaller graphene domains. This was usually observed in the Raman spectra of reduced GO.[4,10,11] In contrast, when our GS were reduced at solvothermal condition (180℃), the reduced GS showed an average D/G intensity ratio lower than that of as-made GS (Fig.2a,b). Since the Raman D/G intensity ratio is proportional to the average size of sp2 domains,[4,12] the solvothermal reduction actually increased the average size of crystalline graphene domains in our GS, which has been unattainable by solution phase reduction of GO.

We observed that the D'/G (Fig.2c) and S3/2D intensity ratios (D' peak is a defect peak due to intra-valley scattering,[12] and S3 peak is a second-order peak due to D-G combination) of the GS also decreased significantly after solvothermal reduction, suggesting that the defect concentration in the GS was much reduced. This was accompanied by a lower oxygen content in the solvothermally reduced GS compared to the as-made GS sample, as revealed by Auger elemental analysis. Although the trend of $I_D/I_G$ (Fig.2b) was not monotonic in reduction temperature, the $I_{D'}/I_G$ (Fig.2c) and $I_{S3}/I_{2D}$ ratios decreased monotonically with increasing reduction temperature. These data suggested effective solvothermal reduction of GS in both oxygen content and defect density.

Electrical devices were fabricated with the as-made and reduced GS. With the increase of reduction temperature, the average two-dimensional (2D) resistivity (defined as R×W/L, where R is resistance of device, W and L are GS width and channel length) of GS decreased (Fig.3b), reaching a minimum down to less than 10kΩ after solvothermal reduction, close to that of pristine peel-off graphene sheets.[13-15] The current-bias ($I_{ds}$-$V_{ds}$) curves of the solvothermally reduced GS were linear, as expected for high quality graphene but not GO. Intrinsic Dirac-like behavior of our solvothermally reduced GS (Fig.3a) was always observed after removal of absorbed molecules by high-current electrical annealing[6,16] in vacuum.

Reduction was also carried out for GO made by a modified Hummers method.[17] As revealed by Raman and electrical measurements, the solvothermally reduced GO had higher quality than GO reduced at lower temperatures. However, unlike decreased D/G intensity ratios of GS with increases in reduction temperature, a monotonic increase of D/G intensity ratio of GO with reduction temperature was observed, indicating large numbers of small $sp^2$ domains exist in GO even after solvothermal reduction. The solvothermally reduced GO showed much broader Raman peaks than the reduced GS, suggesting much more disorder remaining in GO originated from harsh oxidation[12] with a high oxygen content shown by Auger elemental analysis. The average 2D resistivity of solvothermally reduced GO was still more than 100 times higher than that of solvothermally reduced GS. The $I_{ds}$-$V_{ds}$ curves of solvothermally reduced GO showed clear nonlinear exponential characteristics, which fitted well with the 2D variable-range hopping model,[18] suggesting the existence of considerable defective regions between graphene domains. Thus, although our solvothermal reduction is more effective than previous conditions, large numbers of defects still exist in GO that are difficult to reduce, limiting the size of $sp^2$ domains in reduced GO.

The improved effectiveness of our solvothermal reduction could be due to more thorough removal of oxygen functional groups by hydrazine at high temperature, which was corroborated by low oxygen content and decreased "solubility" of the reduced GS. The solvothermal condition could also help to anneal structural defects in GS. However, more work is still needed to reveal the detailed reduction mechanism and pathways.

In conclusion, we developed an effective solvothermal reduction method to lower defects and oxygen content from graphene sheets and GO. We succeeded in decreasing the 2D resistivity of chemically derived graphene to approach that of pristine graphene by the solution phase reduction method. The resulting high-quality GS could be useful for various basic and applied research of graphene materials.

**Figures**

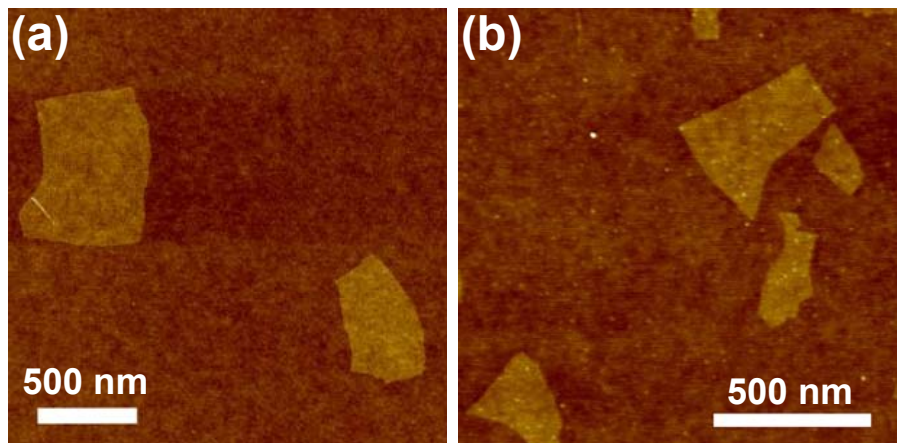

**Fig.1.** AFM images of as-made (a) and solvothermally reduced (b) GS.

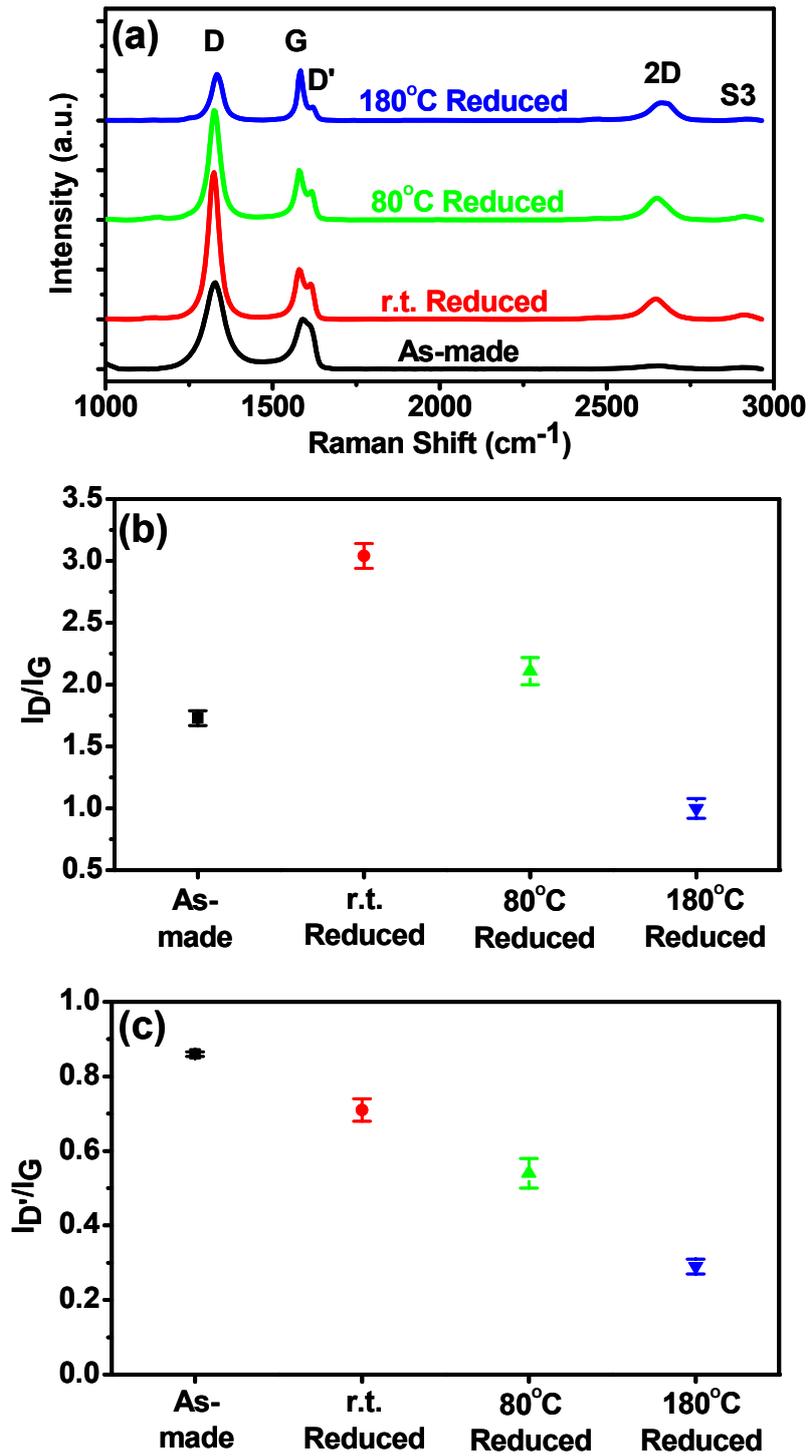

**Fig.2.** Raman characteristics of as-made and different types of reduced GS film samples. (a) Raman spectra, (b) D/G intensity ratio, and (c) D'/G intensity ratio of as-made GS and different types of reduced GS. Error bars are based on Raman spectra taken at 5-10 different spots on each sample.

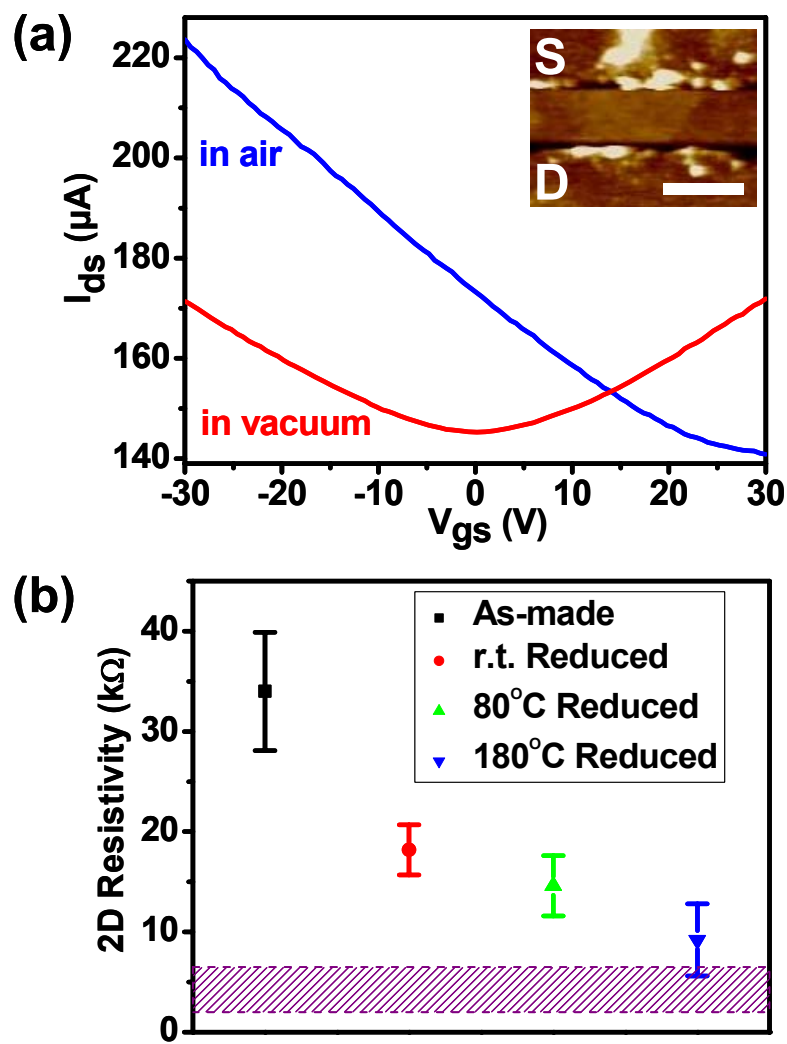

**Fig.3.** Electrical transport characteristics of solvothermally reduced GS. (a) Current-gate voltage ($I_{ds}$-$V_{gs}$) curve of a device. Blue curve: measured in air. Red curve: measured in vacuum after high-current cleaning. Inset shows an AFM image (scale bar: 200nm) of a single GS bridging source (S) and drain (D) electrodes on 100nm $SiO_2$ on p++-Si which is used as a back-gate. (b) 2D resistivity of as-made and different types of reduced GS, with the purple region showing the resistivity range of pristine graphene reported in ref 13-15.